\documentclass[%
reprint,
superscriptaddress,
 amsmath,amssymb,
 aps,
]{revtex4-2}

\usepackage{graphicx}% Include figure files
\usepackage{dcolumn}% Align table columns on decimal point
\usepackage{bm}% bold math
\usepackage{xcolor}
\usepackage[normalem]{ulem}
\usepackage{physics}

\usepackage[pagebackref=false]{hyperref}

\begin{document}

\title{\Impact of competing energy scales on the 
% fourfold 
 shell-filling sequence in elliptic \\ bilayer  graphene quantum dots}

\author{S. M\"oller}
\author{L. Banszerus}
\affiliation{JARA-FIT and 2nd Institute of Physics, RWTH Aachen University, 52074 Aachen, Germany,~EU}%
\affiliation{Peter Gr\"unberg Institute  (PGI-9), Forschungszentrum J\"ulich, 52425 J\"ulich,~Germany,~EU}
\author{A. Knothe}
\affiliation{Institut f\"ur Theoretische Physik, Universit\"at Regensburg, D-93040 Regensburg, Germany}
\author{L. Valerius}
\affiliation{JARA-FIT and 2nd Institute of Physics, RWTH Aachen University, 52074 Aachen, Germany,~EU}%
\author{K. Hecker}
\author{E.~Icking}
\affiliation{JARA-FIT and 2nd Institute of Physics, RWTH Aachen University, 52074 Aachen, Germany,~EU}%
\affiliation{Peter Gr\"unberg Institute  (PGI-9), Forschungszentrum J\"ulich, 52425 J\"ulich,~Germany,~EU}
\author{K.~Watanabe}
\affiliation{Research Center for Functional Materials, 
National Institute for Materials Science, 1-1 Namiki, Tsukuba 305-0044, Japan}
\author{T.~Taniguchi}
\affiliation{International Center for Materials Nanoarchitectonics, 
National Institute for Materials Science,  1-1 Namiki, Tsukuba 305-0044, Japan}%
\author{C. Volk}
\author{C. Stampfer}
\affiliation{JARA-FIT and 2nd Institute of Physics, RWTH Aachen University, 52074 Aachen, Germany,~EU}%
\affiliation{Peter Gr\"unberg Institute  (PGI-9), Forschungszentrum J\"ulich, 52425 J\"ulich,~Germany,~EU}%

\date{\today}


\title{Impact of competing energy scales on the shell-filling sequence in elliptic \\ bilayer  graphene quantum dots}

\begin{abstract} 
We report on a detailed investigation of the shell-filling sequence in electrostatically defined elliptic bilayer graphene quantum dots (QDs) in the regime of low charge carrier occupation, $N \leq 12$, by means of magnetotransport spectroscopy and numerical calculations. 
We show the necessity of including both short-range electron-electron interaction and wavefunction-dependent valley g-factors for understanding the overall fourfold shell-filling sequence. These factors lead to an additional energy splitting at half-filling of each orbital state and different energy shifts in out-of-plane magnetic fields. Analysis of 31 different BLG QDs reveals that both valley g-factor and electron-electron interaction induced energy splitting increase with decreasing QD size, validating theory. However, we find that the electrostatic charging energy of such gate-defined QDs does not correlate consistently with their size, indicating complex electrostatics. These findings offer significant insights for future BLG QD devices and circuit designs.
\end{abstract}

\maketitle

Bilayer graphene (BLG) is a fascinating material as it exhibits a gate-tunable band gap \cite{Min2007Apr, Icking2022Jul, Zhang2009Jun, Oostinga2008Feb}, low spin-orbit interaction \cite{Kurzmann2021OctKondo, Banszerus2021SepSpinOrbit, Banszerus2020MAy, Konschuh2012MarTheoSO}, strong correlation effects \cite{Zhou2022Feb, delaBarrera2022Jul, Seiler2022Aug}, and the possibility to tailor its band structure by proximitizing it with other 2D materials~\cite{Tang2020Apr, Tseng2022Nov, Li2023May, Zhang2014Mar, Ojeda-Aristizabal2009Apr, Gmitra2017Oct, Alsharari2018Jun, Kaloni2014Dec, Island2019Jul}. 
This makes BLG a promising material for future quantum technologies based on 2D materials~\cite{Geim2007Mar, McCann2013Apr, Bertolazzi2013Apr, Zhang2014Jan, Roy2013Nov}.
One way of exploiting the potential of BLG is to confine its charges into quantum dots (QDs) which can serve as building blocks for quantum sensing devices, quantum metrology, spintronics and quantum information units~\cite{Avsar2016Jun, Geim2007Mar, Trauzettel2007Feb, Han2014OctGrapheneSpintronics}.
This is especially true since recent advances in fabrication techniques allow to reliably create ultra clean and highly tunable QDs in BLG by electrostatic soft-confinement, even allowing the creation of QDs with opposite polarity in close proximity~\cite{Tong2021Jan, Banszerus2020Mar, Banszerus2020OctEHcrossover, Banszerus2023MayNature,Banszerus2018Aug, Eich2018Aug, Eich2018Jul, Kurzmann2019Jul}. 
Confining charges in BLG QDs gives rise to fourfold degenerate orbital states in agreement with the spin and valley degrees of freedom present in BLG.
In contrast to QDs in conventional semiconductors, QD states in BLG carry a tunable topological magnetic moment oriented out-of-plane, % of the BLG, 
which is caused by the finite Berry curvature close to the K- and K'-points in gapped BLG \cite{Ren2022May, Knothe2018OctTrigonalWarpingQPC, Moulsdale2020FebKnotheStrain, Knothe2020JunQuartetStates, Park2017, Fuchs2010, McCann2013Apr}.
The orbital magnetic moment, which leads to the valley g-factor, $g_v$,  has opposite sign for the two valleys. It is usually one order of magnitude larger than the spin magnetic moment, thus offering remarkable magnetic field tunability~\cite{Tong2021Jan, Banszerus2021SepSpinOrbit}.
For fully utilizing BLG's high magnetic and electrical tunability, it is therefore of great interest to understand the quantum states and the shell-filling sequence in few electron QDs.
\begin{figure}[!thb]
\centering
\includegraphics[draft=false,keepaspectratio=true,clip,width=\linewidth]{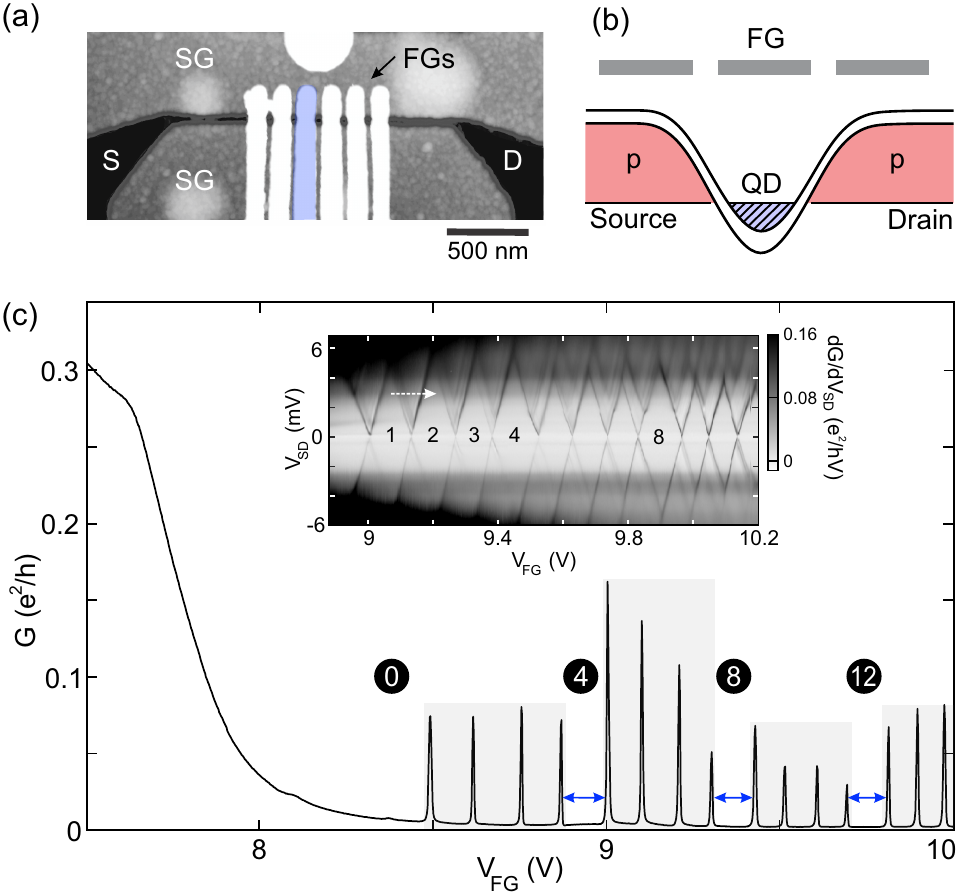}
\caption[Fig01]{%\textbf{BLG QDs created via soft-confinement show fourfold bunching of conductance resonances.}
\textbf{(a)} False color atomic force micrograph of the sample showing a two-layer gate stack with split-gates, source (S) and drain (D), and six 100~nm wide finger gates (FGs). 
\textbf{(b)} Schematic of the band edge profile along the channel illustrating the formation of a QD. 
\textbf{(c)} Conductance through the QD as a function of $V_\mathrm{FG}$ at a bias voltage of $V_\mathrm{SD}=0.3$~mV showing Coulomb peaks. White numbers indicate the electron occupation of the QD while in Coulomb blockade. The addition energy shows fourfold periodicity, highlighted by the blue arrows, which is in agreement with the spin and valley degeneracy of BLG. The inset shows differential conductance through the QD as a function of $V_\mathrm{FG}$ and $V_\mathrm{SD}$.
}
\label{f1}
\end{figure}
Recent works investigate the behavior of orbital states and confirm that Hund's rule is applicable in the framework of non-interacting states for circular BLG QDs~\cite{Eich2018Jul, Garreis2021Apr}. 
\begin{figure}[!thb]
\centering
\includegraphics[draft=false,keepaspectratio=true,clip,width=\linewidth]{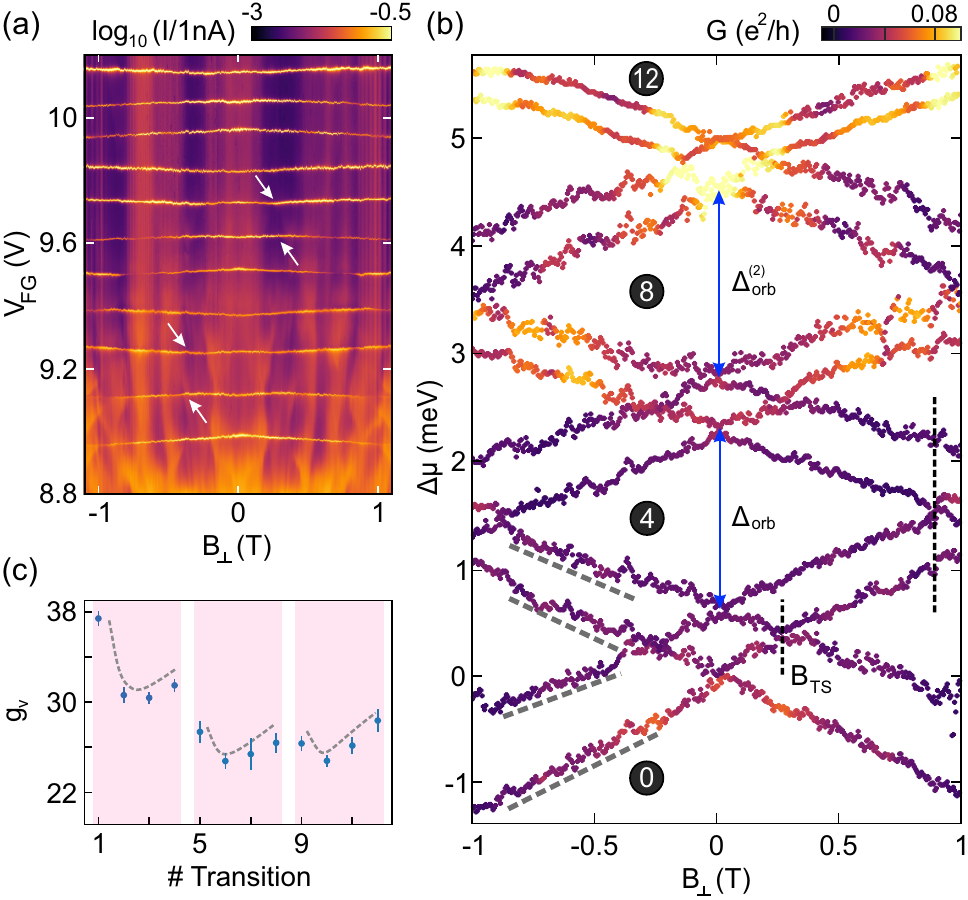}
\caption[Fig02]{%\textbf{Shell-filling of multiple orbitals}
\textbf{(a)} Conductance through the QD at low bias ($V_\mathrm{SD} = 0.2~$mV) as a function of $V_\mathrm{FG}$ and perpendicular magnetic field $B_\perp$. Arrows highlight kinks in the slopes of the peaks, indicating changes of multi-particle ground states.
\textbf{(b)} %Difference in electrochemical potential 
Change in chemical potential, $\Delta\mu$, required to add the next electron to the QD as a function of $B_\perp$, extracted from (a) by subtracting the charging energy $E_\mathrm{C}$. We chose $\Delta \mu = 0$ for the position of the first Coulomb peak at zero magnetic field. White numbers show the electron occupation of the QD in between the Coulomb peaks.
The dashed grey lines indicate where linear fits were performed to determine the valley $g$-factors.
The orbital splitting can be read off from the separations at $B_\perp = 0$, as indicated by the blue arrows, yielding $\Delta_\mathrm{orb} := \Delta^{(1)}_\mathrm{orb} = 1.6 \pm 0.1$~meV for the differences between orbital 1-2 and  similarly $\Delta^{(2)}_\mathrm{orb} = 1.7 \pm 0.1$~meV for the differences between orbital 2-3.
\textbf{(c)} Valley $g$-factors, $g_\nu$, evaluated from the slopes of each charge transition (for more details see {\it Appendix}~\ref{app:g_v_N}). The grey dashed lines are a guide to the eye, highlighting the slight modulation of the g-factor which repeats within each shell.
}
\label{f2}
\end{figure}
However, the question of how short-range electron-electron interactions and different geometries of the confinement potential affect the quantum states remains largely unexplored.

Here we show that for a detailed understanding of the shell-filling sequence of -- at least -- the first three orbital states in elliptic BLG QDs it is crucial to include (i) short-range electron-electron interactions and (ii) wavefunction dependent valley g-factors. The latter leads to a variety of different energy shifts as a function of out-of-plane magnetic fields, while the electron-electron interactions lead to an additional energy splitting at
half-filling of each orbital state.
We also present finite-bias spectroscopy data and investigate the valley g-factor and the short-range electron-electron interaction-induced energy splitting as a function of QD size, expressed in terms of orbital energy splitting. 
By analyzing data from 31 different BLG QD configurations,
we find that both $g_v$ and short-range splitting,  $\delta_2$, increase with decreasing QD size.
Finally, we observe that charging energy, $E_\mathrm{C}$, and orbital splitting, $\Delta_\text{orb}$, show no monotonic dependency. This highlights that the electrostatics of gate-defined BLG QDs, which have low charge carrier density lead regions, are non-trivial, making charging energy not well suited to estimate size differences between different QDs.

To create electrostatically confined QDs, we fabricate heterostructures consisting of a BLG sheet encapsulated within two layers of hexagonal boron nitride~\cite{Engels2014Sep,Wang2013Nov}. The heterostructure is placed on a graphite flake which acts as a back gate (BG)~\cite{Banszerus2018Aug}. Two layers of Cr/Au gates are fabricated on top of the stack: Split gates (SGs) with a separation of $\sim 50~$nm, and 100~nm wide finger gates (FGs) oriented perpendicular to the channel. The two gate layers are separated by a 30~nm thick layer of atomic layer deposited Al$_2$O$_3$. Fig.~\ref{f1}(a) shows an atomic force micrograph of the resulting gate structure. 
By applying voltages of opposite sign to the SGs and the BG we create a perpendicular electric displacement field, which opens up a band gap in BLG and allows to tune the Fermi energy into the band gap~\cite{Icking2022Jul}. This leaves a narrow conducting channel connecting source and drain, which is defined by the SGs. Then, a FG is used to locally invert the polarity of the channel, creating a QD and giving rise to tunnel barriers where the Fermi energy lies within the band gap, as illustrated in Fig.~\ref{f1}(b)~\cite{Eich2018Aug, Banszerus2018Aug}. From the dimensions of SGs and the FG, we expect an elliptical QD with an aspect ratio of $\approx 2$.

The filling sequence of the QD can be studied by transport spectroscopy in a $^3$He/$^4$He dilution refrigerator at a base temperature of around 10~mK using a combination of DC-measurements and standard low-frequency lock-in techniques.
We measure conductance through the channel at low source-drain voltage ($V_\mathrm{SD}$) as a function of applied FG voltage, $V_\mathrm{FG}$, which is shown in Fig.~\ref{f1}(c).
For increasing $V_\mathrm{FG}$, the channel first pinches off ($V_\mathrm{FG} \approx 8.5$ V) before Coulomb resonances appear, indicating the sequential filling of the QD with more than twelve electrons. 
The addition energy of every fourth electron is increased due to the spin and valley degeneracy in BLG, which is highlighted by the blue arrows in Fig.~\ref{f1}(c).
The inset shows the differential conductance $dG/dV_\mathrm{SD}$ as a function of applied bias voltage, where Coulomb diamonds are visible. We can extract the gate lever-arm, $\alpha \approx 0.04$ from the size of the diamonds (see \textit{Appendix}~\ref{app:lever-arm}), allowing to convert $V_\mathrm{FG}$ to chemical potential, $\mu$, according to $\Delta \mu = |e|\,  \alpha  \, \Delta V_\mathrm{FG}$, with the elementary charge $e$.

\begin{figure*}[!thb]
	\centering
\includegraphics[draft=false,keepaspectratio=true,clip,width=\linewidth]{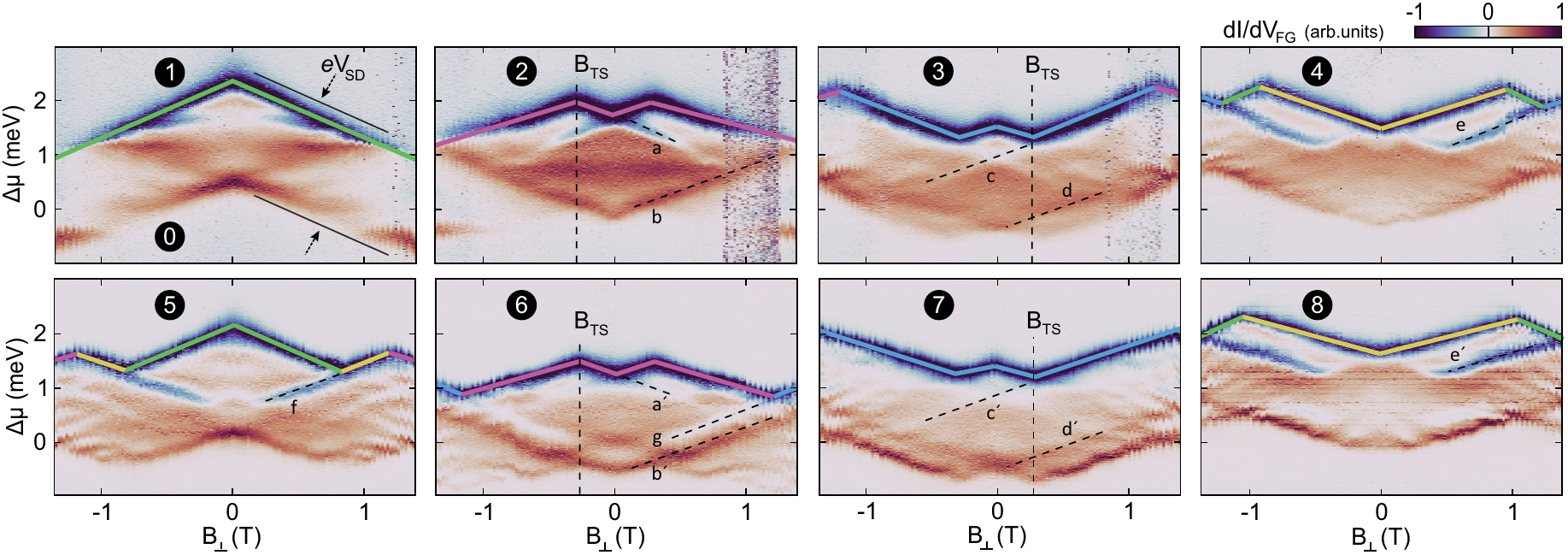}
\caption[Fig03]{%\textbf{Finite bias spectroscopy reveals sequential filling of orbital states.}
Differential transconductance $dI/dV_\mathrm{FG}$ measured across the first eight Coulomb peaks (see white numbers and dashed white arrow in Fig.~\ref{f1}(c)) as a function of $V_\mathrm{FG}$ and $B_\perp$ at V$_\mathrm{SD} = 1.8$~mV. The extent of the conductive region is limited by the bias voltage, as exemplarily shown in the first panel. The colored lines highlight how the ground state (GS) to ground state transition between the $N$ and $N+1$ -- particle state shifts with magnetic field, with the colors corresponding to those in Fig.~\ref{f4}(a). Dashed lines indicate charge transitions involving excited states of the QD, which also show a fourfold pattern, supporting the assumption of sequential filling of orbital states. The features can be explained in terms of transitions between $N$ and $N+1$ particle states of the QD~\cite{Moller2021Dec, Knothe2022AprBatmanTheo}. For example, feature \textbf{b} is a transition from the single particle GS to the valley singlet two particle state $\ket{K^- \uparrow \downarrow} \rightarrow \ket{T^s_{0,\pm} \, S^v}$. It shifts in $B_\perp$ mainly due to the valley Zeeman effect lowering the energy of the single particle GS, while the energy of the valley singlet two-particle state is only slightly altered by the spin Zeeman effect, resulting in a positive slope of the transition. The same is true for the 5- to 6-particle state transition \textbf{b'}, which shows the same behavior as they have the same amount of electrons in the
highest unpaired shell.
A comprehensive discussion of all features is given in \textit{Appendix}~\ref{app:ES_transitions}.}
\label{f3}
\end{figure*}

To investigate the nature of the multi-particle states at each filling, $N$, we perform magnetotransport spectroscopy measurements to probe the spin and valley magnetic moment of each QD state.
Fig.~\ref{f2}(a) depicts conductance through the QD with respect to an out-of-plane magnetic field, $B_\perp$, and the FG voltage, $V_\mathrm{FG}$.
Each Coulomb peak shifts its position on the gate axis as a function of $B_\perp$, featuring kinks that appear simultaneously in adjacent Coulomb peaks, as highlighted by the white arrows. 
Assuming the confinement to be independent of magnetic field, the charging energy corresponds to the minimal distance between two Coulomb peaks~\cite{Eich2018Aug, Garreis2021Apr, Tong2021Jan}. More details about the charging energy and the addition energy are shown in \textit{Appendix}~\ref{app:comparing_energyscales}, Fig.~\ref{energyspacings}.
We subtract the charging energy between neighboring Coulomb peaks and show the result in Fig.~\ref{f2}(b), where we also transformed $V_\mathrm{FG}$ to chemical potential using the lever-arm. 
The y-axis is labeled $\Delta \mu$ to indicate that we subtracted the charging energy and focus only on the change in chemical potential of each Coulomb peak due to the spin and valley Zeeman effect.
For magnetic fields larger than $|B_\perp| \gtrsim 0.3\, \mathrm{T} := B_\mathrm{TS}$ (see black dashed line and $B_\mathrm{TS}$ label in Fig.~2(b)), which we define for later discussions, our observations are consistent with previous works~\cite{Eich2018Aug, Garreis2021Apr, Tong2021Jan} and can be explained by successively filling non-interacting single particle states into the energetically lowest orbital state.
As the valley g-factor is about an order of magnitude larger than the spin g-factor, this gives rise to two positive (adding $\ket{K\uparrow}$ or $\ket{K\downarrow}$) and two negative slopes (adding $\ket{K'\uparrow}$ or $\ket{K'\downarrow}$) per orbital state, according to the magnetic moment of the single particle states added to the QD~\cite{Ihn2010}. 
For large magnetic fields $|B_\perp| \gtrsim 0.85$~T (see vertical dashed black line in Fig.~\ref{f2}(b)), the valley Zeeman splitting becomes larger than the orbital splitting, changing the order in which the shells are filled. 
Thus, the orbital splitting, $\Delta_\mathrm{orb}$, can be directly read off at $B_\perp = 0$, indicated by the blue arrows.
The observed sequential filling of $N = 4-8-12$ (see white numbers) is consistent with numerical calculations of orbital states in BLG QDs (see \textit{Appendix}~\ref{app:calculations}) predicting non-degenerate orbital states for elliptical QDs, which we expect to have due to our gate structure (see Fig.~\ref{f1}(a)).
In contrast, perfectly circular QDs show a $N = 4-12$ filling sequence as they have a degenerate second and third orbital state due to  rotational symmetry~\cite{Garreis2021Apr}.
We evaluate the strength of the valley magnetic moment for each occupation number of the QD by fitting the slopes as indicated by the grey dashed lines in Fig.~\ref{f2}(b)~(see \textit{Appendix}~\ref{app:g_v_N}).
The result is shown in Fig.~\ref{f2}(c), where we plot the valley g-factor, $g_v$, as function of the charge transition, from which it was evaluated.
The g-factors within each shell are nearly constant but exhibit a modulation that repeats after four electrons, see grey dashed lines in Fig.~\ref{f2}(c).
As the valley magnetic moment is determined by the distribution of the electron wavefunction in $k$-space (see examples in Fig.~\ref{fS1}), we expect nearly constant valley g-factors within the same orbital state.
We speculate that the modulation within each shell arises from slight changes in the electrostatics of the QD due to the higher FG voltage and the additional electron occupying the same shell (see \textit{Appendix}~\ref{app:g_v_N}).
This modulation is pronounced strongest between the first and second electron entering the QD, which is expected, as there the QD is just formed after pinching off the channel and the confinement potential initially changes significantly with increasing $V_\mathrm{FG}$~\footnote{This can also be seen from the charging energy of the QD, which decreases strongly for the first few charge carriers entering the QD and then stabilizes, see Refs. \cite{Banszerus2020Mar, Eich2018Aug}.}\cite{Banszerus2020Mar, Eich2018Aug}.
Nevertheless, as the modulation of g-factors also shows a fourfold pattern, the observation is consistent with the assumption that only one orbital state is filled at a time.

This assumption is further supported by finite bias spectroscopy, which allows to probe both ground and excited states of each charge transition.
We measure across the first eight Coulomb peaks at $V_\mathrm{SD} = 1.8$~mV, as exemplarily illustrated by the white dotted arrow in the inset in Fig.~\ref{f1}(c). 
The results are displayed in Fig.~\ref{f3}, where we plot differential transconductance $dI/dV_\mathrm{FG}$ as a function of $B_\perp$ and $\Delta \mu$, where the latter was obtained from $V_\mathrm{FG}$ using the lever-arm.
White numbers indicate the charge occupation of the QD in the Coulomb blockaded region.
The outline of the conducting region (see black dashed arrows in the upper most left panel in Fig.~\ref{f3}) is determined by the size of the bias window, $V_\mathrm{SD}$ and shifts with $B_\perp$ exactly as the charge resonances in Fig.~\ref{f2}(a,b)~\cite{Moller2021Dec}.
Within the conducting region, we can observe additional features which originate from excited state transitions of each charge transition, increasing or decreasing the tunnel current as they enter or leave the bias window. 
A detailed state assignment and more information on how to interpret the finite bias data are found in \textit{Appendix}~\ref{app:ES_transitions} and Refs.~\cite{Moller2021Dec, Knothe2022AprBatmanTheo}.
Here, it is important to note that similar features of excited state transitions are visible at corresponding charge transitions, i.e. 1-5, 2-6, 3-7, 4-8, as can be seen by comparing the upper and lower panels, as highlighted by the labeled dashed lines.
Thus, the fourfold pattern is also present here, further supporting the assumption that only one orbital state is filled at a time for $B_\perp < 0.85$~T.

For magnetic fields $|B_\perp| < B_\mathrm{TS} \approx 0.3$~T, we observe an additional energy scale, which causes a two-fold bunching of Coulomb resonances within each orbital (see $B_\mathrm{TS}$ in Fig.~\ref{f2}(b) and Fig.~\ref{f3}). 
This is a robust feature present in all works where few charge carrier gate-defined QD states in BLG are investigated~\cite{Eich2018Jul, Kurzmann2019Jul, Kurzmann2021OctKondo, Garreis2021Apr, Tong2021Jan, Moller2021Dec, Banszerus2020Mar}, strongly indicating that electron-electron interaction needs to be taken into account for understanding the filling of individual shells.

\begin{figure}[!t]
\centering
\includegraphics[draft=false,keepaspectratio=true,clip,width=\linewidth]{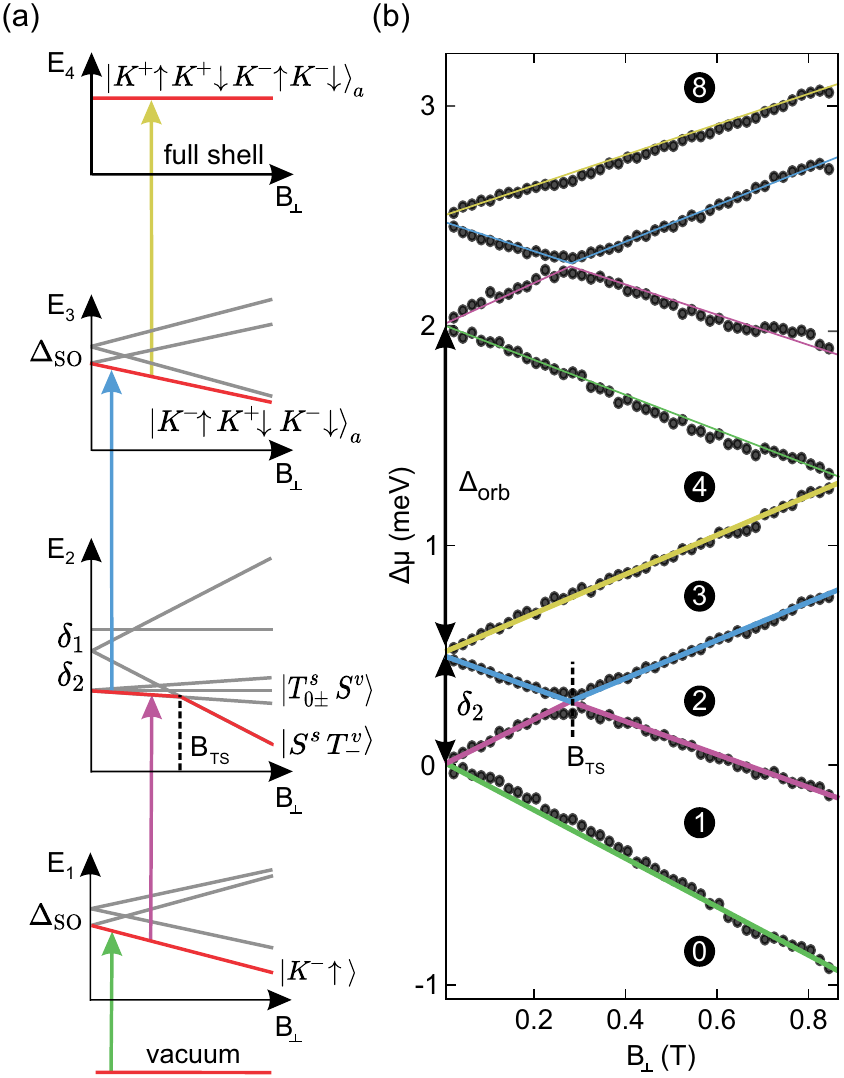}
\caption[Fig04]{%\textbf{Filling of single orbital shells shows electron-electron interactions at half filling.}
\textbf{(a)} Energy dispersion of the first four multi-particle states in a perpendicular magnetic field. The ground state is highlighted in red, while colored arrows correspond to the chemical potential required to add the next electron onto the QD. At $B_\mathrm{TS}$ the GS of the two particle states changes from a spin triplet -- valley singlet to a spin singlet -- valley triplet. 
\textbf{(b)} Change in chemical potential, $\Delta\mu$, required to add the next electron to the QD as a function of $B_\perp$, as in Fig.~\ref{f2}(b) but extracted from Fig.~\ref{f3}. White numbers indicate the occupation of the QD when in Coulomb blockade. The colored lines correspond to the length of the arrows in (a). Orbital splitting, $\Delta_\mathrm{orb}$, and electron-electron interaction strength, $\delta_2$, can directly be read off.
}
\label{f4}
\end{figure}

To understand the energy scales involved when filling one orbital state, we go beyond the single particle model and take a closer look at involved multi-particle states.
Fig.~\ref{f4}(a) shows the energy of the first four multi-particle states, $E_N$ with $N = 1,...,4$, as a function of $B_\perp$.
Note that constant energy differences between states with different occupation number, $N$, are not displayed.
For one electron in the QD, there are four single particle states available, which shift due to the spin and valley Zeeman effect and which are experiencing a small spin-orbit splitting, $\Delta_\mathrm{SO} \approx 60-80$~$\mu$eV~\cite{Kurzmann2021OctKondo, Banszerus2021SepSpinOrbit, Konschuh2012MarTheoSO}.
The single particle ground state (GS) shifts therefore according to
\begin{equation}
   E^\mathrm{GS}_1  = - \frac{1}{2} (g_s + g^{(1)}_v ) \mu_B B_\perp ,
\end{equation}
with the Bohr-magneton, $\mu_B$, the spin g-factor, $g_s$, and the valley g-factor, $g^{(1)}_v$, where the upper index ($N$) refers to the occupation number $N$ of the QD, allowing to account for the slight occupation number dependency of the valley g-factor~\cite{Moller2021Dec}.
As we assume only one orbital state to be filled at a time, the orbital part of the multi-particle wavefunction is symmetric. The Pauli-principle then requires the spin and valley part of the multi-particle wavefunction to be anti-symmetric, strongly reducing the available QD states at each filling. For the case of $N = 2$, there are thus only six two particle states available~\cite{Moller2021Dec}. 
These are grouped into three valley triplet -- spin singlet states, $\ket{S^s \,\, T^v_{0,\pm}}$, and three spin triplet -- valley singlet states, $\ket{T^s_{0,\pm}\,\, S^v}$, which are separated by $\delta_{1}$ and $\delta_{2}$ due to short-range electron-electron interaction~\cite{Moller2021Dec, Knothe2022AprBatmanTheo, Knothe2020JunQuartetStates, Lemonik2010Nov, Lemonik2012Jun}. 
This leads to the interesting situation where the two particle GS is a spin triplet at low magnetic fields but becomes a spin singlet at $B_\perp := B_\mathrm{TS}$, indicated by the index 'TS'. The change in GS occurs as soon as the valley Zeeman effect of $\ket{S^s \,\, T^v_{-}}$ compensates the short-range splitting $\delta_2$.
Consequently, the magnetic field dependency of the two particle ground state has two different slopes, 
\begin{align}
    E^\mathrm{GS}_2 =\begin{cases} - g_s \mu_B B_\perp &  \text{for} \;\;\; B < B_\mathrm{TS}, \\
          - g^{(2)}_v  \mu_B B_\perp   &  \text{for} \;\;\; B > B_\mathrm{TS}.
       \end{cases}
\end{align}
Still assuming sequential filling of orbitals, for $N = 3$, there are only four available states. 
This is due to the fact all three electrons are occupying the same orbital state and thus the Pauli-principle requires them to have different spin and valley quantum numbers, prohibiting fully valley or spin polarized three particle states.
\begin{figure}[!h]
\centering
\includegraphics[draft=false,keepaspectratio=true,clip,width=0.995\linewidth]{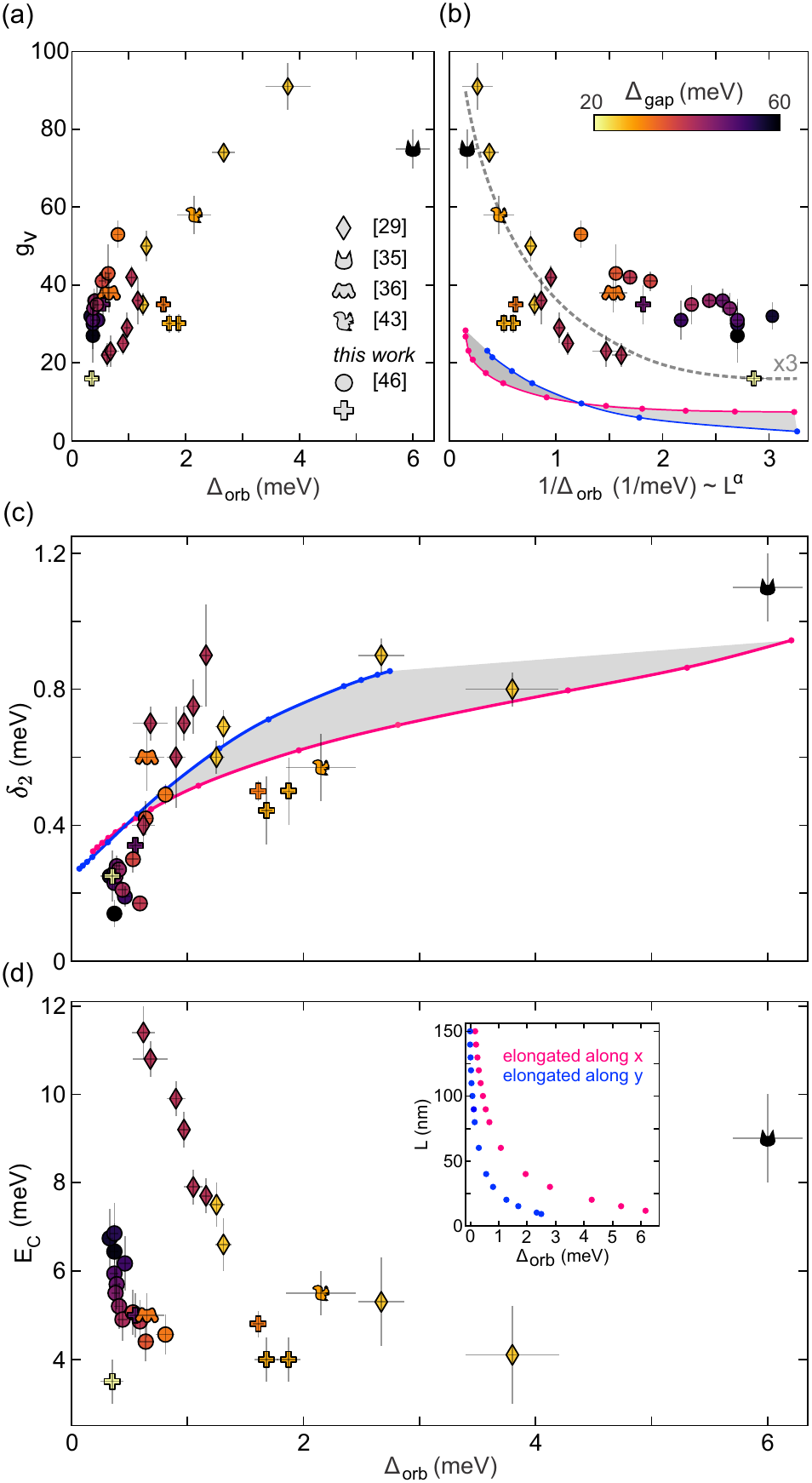}
\caption[Fig01]{%\textbf{Geometry of the QD critically influences the energyscales of the shell-filling.}
\textbf{(a)} Valley g-factor of the first shell in single QDs in literature \cite{Eich2018Jul, Tong2021Jan, Kurzmann2019Jul, Garreis2021Apr} and in our work as a function of orbital splitting. The colorscale of the markers indicates the size of the band gap used to confine each QD.
\textbf{(b)} shows the same data as in (a) but as a function of $1/\Delta_\mathrm{orb}$, which is proportional to the length parameter of the QD, $L^\alpha$, with $\alpha \approx 2$. The colored lines indicate the strength of the valley g-factor obtained by numerical calculations of elliptical QDs, oriented along the x- (pink) and y-axis (blue) of the BLG lattice. The dashed line is a guide to the eye, illustrating how the experimental data is a factor of $\approx 3$ larger compared to the theoretic expectation. 
\textbf{(c)} Strength of the electron-electron interaction, $\delta_2$, as a function of orbital splitting. Again, the colored lines are obtained by numerical calculations of elliptical QDs.
\textbf{(d)}~Charging energy, $E_\mathrm{C}$,  as a function of orbital splitting $\Delta_\mathrm{orb}$. The inset shows the calculated $\Delta_\mathrm{orb}$ between the first and second single particle orbital for different length parameters, $L$, of the different BLG QDs (see {\it Appendix}~\ref{app:calculations}).
}
\label{f5}
\end{figure}
Therefore the three particle GS, $\ket{K^- \uparrow \, K^+ \downarrow \, K^-\downarrow}_a$, with the $a$ indicating this state to be an anti-symmetric superposition of $\ket{K^-\uparrow}, \ket{K^+\downarrow}, \ket{K^-\downarrow}$, has a similar magnetic field dispersion as the single particle GS, resulting in
\begin{equation}
   E^\mathrm{GS}_3  = - \frac{1}{2} (g^{(3)}_v - g_s ) \mu_B B_\perp .
\end{equation}
For a full shell, $N = 4$ there is only one state available, which does not shift in $B_\perp$, as that state contains electrons with all four possible spin and valley combinations, leading to zero total magnetic moment and  
\begin{equation}
     E^\mathrm{GS}_4  = const.
\end{equation}
The chemical potential of each charge transition is given by $\mu_{N} = E_{N} - E_{N - 1}$, their magnetic field dispersion consequently depends on both involved multi-particle states. 
This is illustrated by the colored arrows in Fig.~\ref{f4}(a), whose lengths correspond to the position of each Coulomb peak on the $V_\mathrm{FG}$ axis, i.e. chemical potential axis.
Fig.~\ref{f4}(b) shows the change of chemical potential of the first eight Coulomb peaks as a function of $B_\perp$, similar to Fig.~\ref{f2}(b) but evaluated from the data of Fig.~\ref{f3} (see colored lines) for a better signal-to-noise ratio.
Comparing the color code in Fig.~\ref{f4}(a) and 4(b), we can understand where the two-fold bunching of Coulomb peaks within one shell comes from.
For low magnetic fields, the two particle GS is a valley singlet, shifting only due to the spin Zeeman effect, while the single particle GS shifts due to the valley and spin Zeeman effect, resulting in a positive slope for the second Coulomb peak, $\mu_2 (B_\perp < B_\mathrm{TS}) = E_2 - E_1 = \frac{1}{2} (g^{(1)}_v - g_s) \mu_B B_\perp > 0 $. 
Only for large magnetic field, $B_\perp > B_\mathrm{TS}$, where the two particle GS becomes a valley triplet, the slope becomes negative.
In this regime the second Coulomb peak shifts with $\mu_2 (B_\perp > B_\mathrm{TS}) = - \frac{1}{2} (2 g^{(2)}_v - g^{(1)}_v - g_s) \mu_B B_\perp < 0 $. Neglecting the slight modulation of valley g-factors and assuming $g^{(2)}_v \approx g^{(1)}_v$,  we recover the result one would expect when assuming non-interacting single particle states.
As the three particle states behave like single particle states in $B_\perp$, the third Coulomb peak mirrors the behavior of the second one, and similarly, the forth Coulomb peak mirrors the first one.
This behavior is illustrated by the blue and yellow arrows in Fig.~\ref{f4}(a).
Note that the data is not perfectly symmetric within each shell as the valley g-factor sensibly depends on the shape of the confinement potential, which changes slightly for increasing charge carrier occupation~\cite{Garreis2021Apr}, as also shown in Fig.~\ref{f2}(c).
Summing up, it is solely the change of the two particle GS at $B_\mathrm{TS}$ due to the short-range electron-electron interaction which causes the two-fold bunching of Coulomb peaks, and one can directly read off its strength $\delta_2$ from the data in Fig.~\ref{f4}(b) (and Fig.~\ref{f2}(b)).

Utilizing this understanding, we investigate how different spatial dimensions influence the energy scales determining the shell-filling in BLG QDs.
In total, we gathered 31 data points from QDs in different electrostatic environments, both from our samples and from the work of the Ensslin group \cite{Moller2021Dec, Tong2021Jan, Eich2018Jul, Kurzmann2019Jul, Garreis2021Apr, ETH_Data_repository}. 
Together, the strength of the valley g-factor and electron-electron interaction govern the shell-filling at low magnetic field, because they determine the magnetic field $B_\mathrm{TS}$, at which the valley Zeeman splitting compensates the short-range interaction and the two particle GS changes from $\ket{T^s_{-} \, S^v}$ to $\ket{S^s \, T^v_{-}}$,
\begin{equation}
    B_\mathrm{TS} = \frac{\delta_2}{(g^{(2)}_v - g_s) \mu_\mathrm{B}}.
\end{equation}
Meanwhile, for high out-of-plane magnetic fields, it is the orbital splitting and the valley Zeeman effect dominating the shell-filling. 
In Fig.~\ref{f5}(a,b), we show the strength of the valley g-factor of the first orbital as a function of $\Delta_\mathrm{orb}$ and $1/\Delta_\mathrm{orb}$, with the latter being proportional to $L^\alpha$, with the length (size) parameter of the QD, $L$, and $\alpha \approx 2$, depending on the shape of the confining potential (see \textit{Appendix}~\ref{app:calculations}). 
The color scale of the markers corresponds to the magnitude of the band gap, $\Delta_\mathrm{gap}$, opened in BLG to confine each respective QD, which was estimated from the applied electric displacement field following Ref.~\cite{Icking2022Jul}.
While the exact value of the valley magnetic moment depends sensibly on the shape of the confinement potential of the QD, its orientation with respect to the BLG lattice, the strain in the system, and the magnitude of the band gap~\cite{Tong2021Jan, Knothe2020JunQuartetStates, Knothe2018OctTrigonalWarpingQPC, Moulsdale2020FebKnotheStrain}, we still observe a clear trend towards larger valley g-factors for larger orbital splittings, i.e. for smaller QDs.
This trend is consistent with numerical calculations of elliptical QDs, which are included in Fig.~\ref{f5}(b), with the (blue) pink line corresponding to the QD being oriented along the (y-) x-axis of the BLG lattice. 
The model adds two spatially varying functions, describing the band gap opening and the confinement, to the four-band Hamiltonian of BLG~\cite{McCann2013Apr} and diagonalizes it in order to obtain the orbital wavefunctions of the QD states.
We integrate the Berry curvature induced orbital magnetic moment over the distribution of the first orbital wavefunction in $k$-space in order to obtain the valley g-factor~\cite{Knothe2020JunQuartetStates}.
However, the magnitude of the calculated valley g-factors is too small by a factor of $\sim 3$ compared to the experimental data in Fig.~\ref{f5}(b).
%
%We speculate that the van-der-Waals stacking technique used to fabricate the samples often introduces strain to the BLG, which causes a significant enhancement of the valley magnetic moment~\cite{Moulsdale2020FebKnotheStrain}.
%
It is likely that the van-der-Waals stacking technique used to fabricate the samples introduces strain to the BLG, which is not included in our calculation. However, it has been shown that already little strain can cause significant enhancement of the valley magnetic moment~\cite{Moulsdale2020FebKnotheStrain, Lee2020Mar}. 
As one would expect, the strength of the short-range electron-electron interaction increases for larger orbital splittings \cite{Knothe2020JunQuartetStates}, i.e. smaller QDs, which can be seen in Fig.~\ref{f5}(c), where we plot $\delta_2$ as a function of $\Delta_\mathrm{orb}$. 
This behavior is reproduced by numerical calculations using the same model and parameters as in Fig.~\ref{f5}(a,b). 
More details about the model can be found in \textit{Appendix}~\ref{app:calculations}.

Furthermore, we investigate the dependency between charging energy, $E_\mathrm{C}$, and orbital splitting, $\Delta_\mathrm{orb}$, which is shown in Fig.~\ref{f5}(d). Interestingly, these two quantities have no simple functional relation. For the same $\Delta_\mathrm{orb}$, $E_\mathrm{C}$ may vary by more than a factor of 3, and similarly, for the same $E_\mathrm{C}$, $\Delta_\mathrm{orb}$ may vary by more than a factor of 5. As $E_\mathrm{C}$ depends on the capacitive coupling of the QD to its surrounding, according to $E_\mathrm{C} = e^2 / C_\Sigma$, with the total capacitance of the QD, $C_\Sigma$ \cite{Ihn2010}, it becomes clear that the investigated QDs were confined in distinctly different electrostatic environments. This is not surprising as they were confined in samples with different thicknesses of the hBN and Al$_2$O$_3$ dielectrica, different gate geometries and different gate voltages. Together with the fact that the QD leads are usually in a regime of low charge carrier density, where quantum capacitance effects are relevant, it becomes clear that the electrostatics of gate-defined BLG QDs are far from trivial. The inset of Fig.~\ref{f5}(d) shows $\Delta_\mathrm{orb}$ calculated for different sizes of the QD, quantified by the length parameter, $L$. As expected, there is a monotonic decrease of  $\Delta_\mathrm{orb}$ for larger QDs. Together with the non-trivial dependency between $E_\mathrm{C}$ and $\Delta_\mathrm{orb}$, it is apparent that charging energy is in general not well suited to estimate QD size across different QDs.

%%%%%%%%%%%%%%%%%%%%%%%%%%%%%%%%%%%%%%%%% Summary %%%%%%%%%%%%%%%%%%%%%%%%%%%%%%%%%%%%%%%%%
In summary, we provide a full understanding of the multi-particle spectrum and the fourfold shell-filling sequence of BLG QDs.
In particular, we have shown that the short-range electron-electron interaction, the orbital energy and the (state-dependent) valley $g$-factor are of particular relevance and are critically influenced by the size of the QD.
For low magnetic fields, the valley magnetic moment and the electron-electron interaction strength determine the shell-filling, leading to a spin-triplet two particle ground state and filling according to Hund's rule. For high magnetic fields, shell-filling is dominated by the strong valley magnetic moment, which will eventually even change the order in which orbital states are filled.
Understanding the shell-filling sequence enables future works on few-charge carrier multi-QDs in BLG. Foremost, this is important for evaluating the possiblity of spin-, valley-  and Kramer's qubits in BLG and in which regime of single or multi-QDs they can be operated. They can also be used to probe the spin and valley configuration of correlated phases in BLG. In addition, creating QDs in proximitized BLG can allow to sensibly quantify the influence of the functional layer on the band structure of BLG. Additionally, our detailed quantitative analysis of the involved energy scales provides valuable insights for designing future BLG QD devices and circuits.

\textbf{Acknowledgements} The authors thank S.~Trellenkamp, F.~Lentz and M. Otto for their support in device fabrication, as well as V. Fal'ko for his enlightening contributions to our discussions.
This project has received funding from the European Union's Horizon 2020 research and innovation programme under grant agreement No. 881603 (Graphene Flagship) and from the European Research Council (ERC) under grant agreement No. 820254, the Deutsche Forschungsgemeinschaft (DFG, German Research Foundation) under Germany's Excellence Strategy - Cluster of Excellence Matter and Light for Quantum Computing (ML4Q) EXC 2004/1 - 390534769, and by the Helmholtz Nano Facility~\cite{Albrecht2017May}. K.W and T.T acknowledge support from the JSPS KAKENHI (Grant Numbers 19H05790 and 20H00354).

\section*{Appendix}
\subsection{Occupation dependency of the lever-arm}
\label{app:lever-arm}
\begin{figure}[h]
\centering
\includegraphics[draft=false,keepaspectratio=true,clip,width=0.9\linewidth]{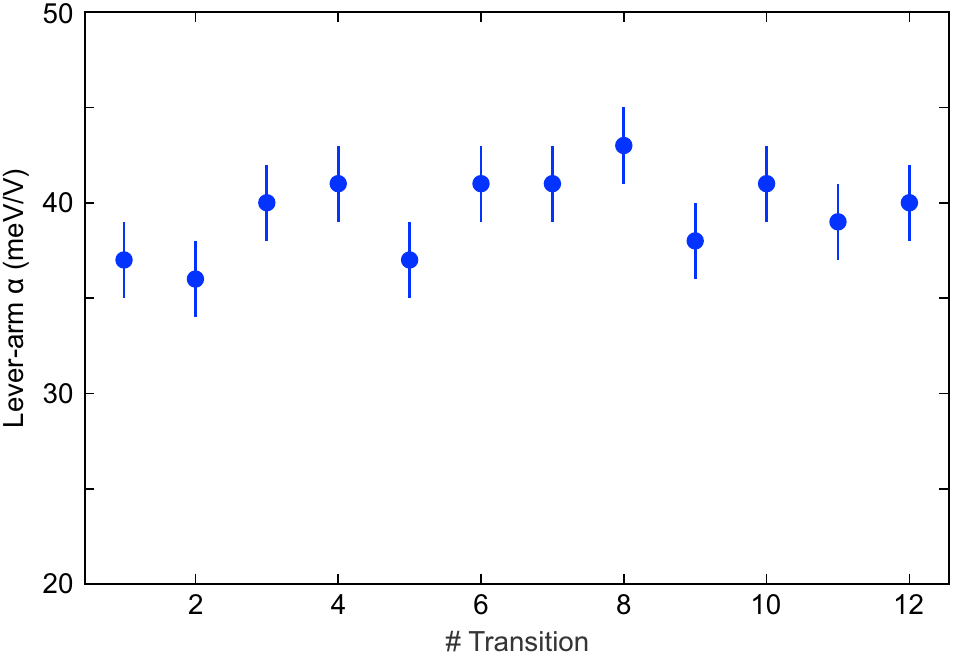}
\caption[Fig.]{
Variation of the FG lever-arm $\alpha$ with respect to charge transition. } 
\label{lever_arm}
\end{figure}

We have extracted the lever-arms from the outline of the conducting region of the finite bias spectroscopy data presented in the inset of Fig.~\ref{f1}(c) as well as from the data in Fig.~\ref{f3}. The lever-arm varies between 36 and 43 meV/V for different charge transitions, with an estimated uncertainty of $\approx \pm 2$~meV/V, as displayed in Fig.~\ref{lever_arm}.

\subsection{Comparison between  $E_\mathrm{add}$, $E_\mathrm{C}$, $\Delta_\mathrm{orb}$ and  $\delta_2$}
\label{app:comparing_energyscales}
\begin{figure}[h]
\centering
\includegraphics[draft=false,keepaspectratio=true,clip,width=0.8\linewidth]{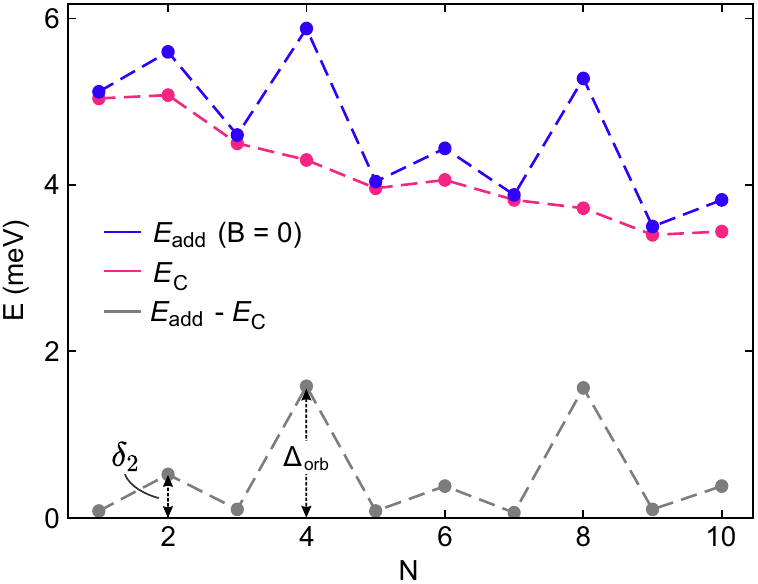}
\caption[Fig01]{Plotting charging energy, $E_\mathrm{C}$, addition energy, $E_\mathrm{add}$, and their difference at zero magnetic field  as a function of QD occupation, $N$. In the latter the influence of the short-range electron-electron interaction, $\delta_2$, and of the orbital splitting $\Delta_\mathrm{orb}$ is visible.}
\label{energyspacings}
\end{figure}

From the magnetotransport data presented in Fig.~\ref{f2}(a), we extract the addition energy, $E_\mathrm{add}$ at zero magnetic field, and the charging energy, $E_\mathrm{C}$, which corresponds approximately to the minimal distance between neighboring Coulomb peaks~\cite{Eich2018Aug, Ihn2010}. We observe a monotonic decrease of $E_\mathrm{C}$ for higher occupation numbers and see the influence of the orbital splitting, $\Delta_\mathrm{orb}$, and the short-range splitting, $\delta_2$, on the addition energy, as depicted in Fig.~\ref{energyspacings}.

\subsection{Numerical calculations of elliptical QDs}
\label{app:calculations}
For calculating the blue and pink curve in Fig.~\ref{f5}(b,c), we follow the approach of Refs.~\cite{ Knothe2020JunQuartetStates, Moller2021Dec, Knothe2022AprBatmanTheo, Tong2021Jan}. We describe single electrons in BLG QDs using the four-band Hamiltonian \cite{McCann2013Apr}, including a spatially varying confinement potential, $ U(\mathbf{r})$, and band gap profile, $\Delta(\mathbf{r})$,
\begin{align}
 \text{H}_{\xi}\!=\! \!&
\setlength{\arraycolsep}{+5pt} \begin{pmatrix} 
 U -\xi\frac{1}{2}\Delta  & \xi v_3\pi & 0 &\xi v \pi^{\dagger}\\
\xi v_3 \pi^{\dagger}&  U +\xi\frac{1}{2}\Delta  & \xi v\pi &0\\
 0 &  \xi v\pi^{\dagger} &   U +\xi\frac{1}{2}\Delta  &   \gamma_1\\
\xi v\pi & 0 &   \gamma_1 &  U -\xi\frac{1}{2}\Delta 
\end{pmatrix},
\label{eqn:H}
\end{align}
in the two valleys $K^{\xi}$ labeled by the valley index $\xi=\pm 1$. The Hamiltonian in Eq.~(\ref{eqn:H}) is written in the Bloch basis $\psi_{K^+}=(\psi_{A},\psi_{B^{\prime}},\psi_{A^{\prime}},\psi_{B})$ in valley $K^+$, and $\psi_{K^-}=(\psi_{B^{\prime}},\psi_{A},\psi_{B},\psi_{A^{\prime}})$ in valley $K^-$  (with electron's amplitudes on the BLG sublattices  $A$ and $B$ in the top, and  $A^{\prime}$ and $B^{\prime}$ in the bottom layer) and in terms of the momenta $\pi=p_x+ip_y,\,  \pi^{\dagger}=p_x-ip_y$, and parameters $v=1.02 \cdot 10^6 \text{ m/s}$,  $v_3\approx0.12 v$, and  $\gamma_1\approx0.38\text{ eV}$.

We include the influence of the electrostatic gates by choosing the confinement potential and gap profile to model the experiment,
\begin{align}
    U(\mathbf{r}) = U_0 /  &\mathrm{cosh} \left( \frac{\sqrt{(\frac{x}{a})^2 + (\frac{y}{b})^2}}{L} \right),\, \mathrm{and} \\
    \Delta(\mathbf{r}) = \Delta_{0} -  \Delta_{mod} / &\mathrm{cosh} \left( \frac{\sqrt{(\frac{x}{a})^2 + (\frac{y}{b})^2}}{L} \right),
\end{align}
with the band-gap opened by SG and BG of$\Delta_{0} = 35\,$meV, the maximal FG induced gap modulation $\Delta_{mod}=0.15 \Delta_{0}$, the maximal potential depth $U_0 = 15\,$meV, an ellipticity of $\frac{a}{b} = 2 \, \, (\frac{1}{2})$ for the two perpendicular orientations of the elliptical QD along the BLG lattice, and the width parameter $L$. 

\begin{figure}[b]
\centering
\includegraphics[draft=false,keepaspectratio=true,clip,width=\linewidth]{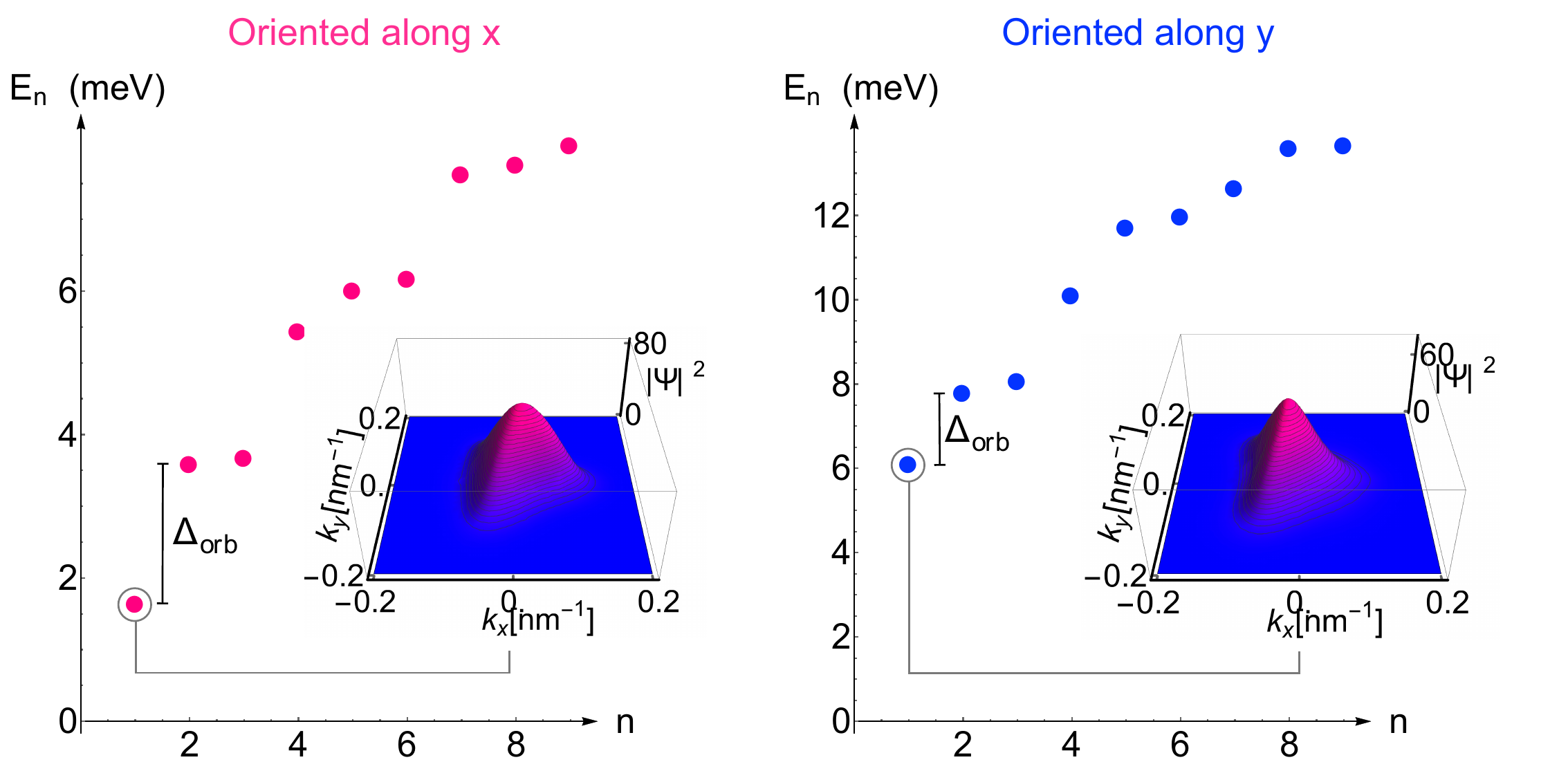}
\caption[Fig01]{Single particle orbital states, $n$, extracted from diagonalising Eq.~\eqref{eqn:H}: We exemplify $\Delta_\mathrm{orb}$ and the ground state wave function in momentum space for the QD widths $L=40$~nm (oriented along the $x$-direction, left) and $L=15$~nm (oriented along the $y$-direction, right), which lead to comparable orbital splittings $\Delta_\mathrm{orb}\approx 2$~meV. The second and third orbital state are not fully degenerate due to ellipticity.}
\label{fS1}
\end{figure}

We numerically diagonalize the four-band Hamiltonian in Eq.~(\ref{eqn:H}) \cite{Knothe2018OctTrigonalWarpingQPC, Knothe2020JunQuartetStates} to obtain the single particle orbital states for the confining QD potential widths, $L = 10 - 150\,$~nm, see Fig.~\ref{fS1}. For each value of L, we extract the energy difference between the ground and the first excited orbital state, $\Delta_\mathrm{orb}$, and estimate the topological valley g-factor, $g^{(1)}_v$ of the lowest orbital state by calculating how much orbital magnetic moment, $M_z$, is picked up by the orbital ground states wave function, $\Psi$, in momentum space,
valley g-factor\cite{Lee2020Mar, Knothe2020JunQuartetStates, Tong2021Jan}, 
\begin{equation}
\mu_B{g}_v=\int d\boldsymbol{k} M_z(\boldsymbol{k} )|\Psi(\boldsymbol{k} )|^2.
\end{equation}
Here, the orbital magnetic moment, induced by the non-trivial Berry curvature,  of BLG's Bloch bands, is defined as $\mathbf{M}(\mathbf{k})=M_z(\mathbf{k})\mathbf{e}_z$,   \cite{Park2017, Fuchs2010, Xiao2010, Chang1996a} with
\begin{equation}
{M}_z=-i\frac{e}{2\hbar}\langle\mathbf{\nabla}_{\mathbf{k}}\Phi(\mathbf{k})|\times [\epsilon(\mathbf{k})-H(\mathbf{k})] |\mathbf{\nabla}_{\mathbf{k}}\Phi(\mathbf{k})\rangle\cdot\mathbf{e}_z,
\label{eqn:Berry}
\end{equation}
where, $\mathbf{\nabla}_{\mathbf{k}}=(\partial_{k_x},\partial_{k_y})$, $"\times"$ is the cross product,  $\epsilon(\mathbf{k})$ is the band energy, and $\Phi$ is the corresponding Bloch state.

\begin{figure}[h]
\centering
\includegraphics[draft=false,keepaspectratio=true,clip,width=0.85\linewidth]{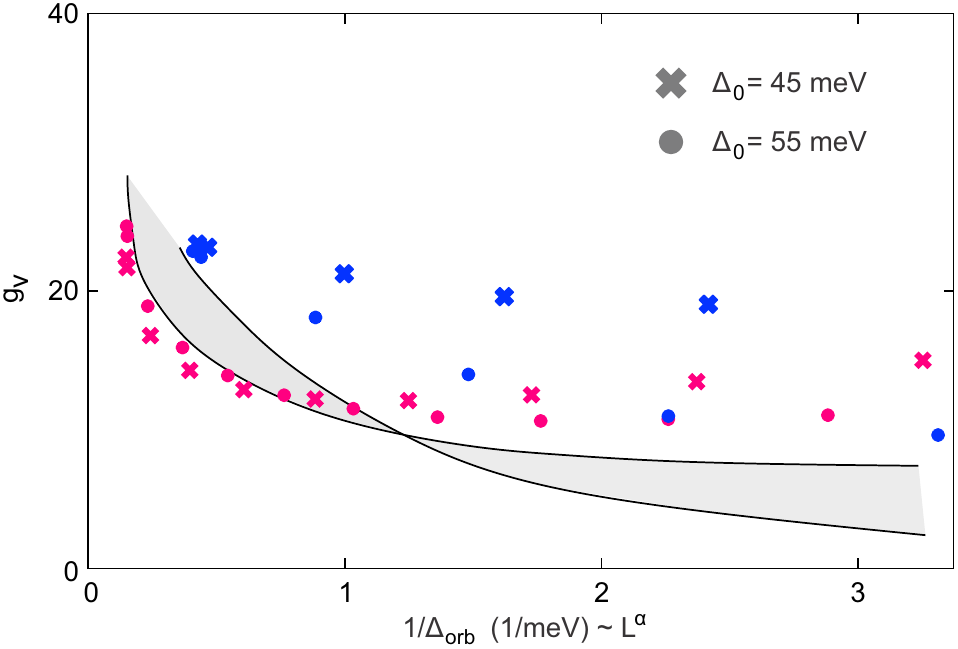}
\caption[Fig.]{Valley g-factor of the first single particle orbital state a function of $1/\Delta_\mathrm{orb}$, which is proportional to the length parameter of the QD, $L^\alpha$, with $\alpha \approx 2$. The black lines correspond to the calculated values for $\Delta_0 = 35$ meV, as is shown in the main manuscript in Fig.~5(b). The cross (round) data points correspond to $\Delta_0 = 45 \, (55)$ meV, while the pink (blue) markers refer to the QD being oriented along the x (y) axis of the BLG lattice. } 
\label{g_v_diff_gaps}
\end{figure}

Also, we have performed the calculations for higher band-gaps, $\Delta_0 = 45 \, (55)\,$meV, as defined in Eq.~(8). The results are shown in Fig.~\ref{g_v_diff_gaps}, where we plot $g_v$ of the first single particle orbital state a function of $1/\Delta_\mathrm{orb}$. The black lines correspond to the calculated values for $\Delta_0 = 35$ meV, as is shown in the main manuscript in Fig.~\ref{f5}(d). The cross (round) data points correspond to $\Delta_0 = 45 \, (55)$ meV, while the pink (blue) markers refer to the QD being oriented along the x (y) axis of the BLG lattice. For very small QDs, it shows that the magnitude of the valley g-factor is reduced for larger band-gaps compared to smaller ones, which can be observed for the data point at $\Delta_\mathrm{orb} = 6\,$meV in Fig.~\ref{f5}(b). Contrarily, for large QDs, smaller band-gaps actually lead to slightly smaller valley g-factors, which can be observed for the data point with $g_v = 18$ in Fig.~\ref{f5}(b). 
However, the trend of smaller valley g-factors for larger QDs remains the same, independent of the chosen band-gap. The same is true for the discrepancy between simulation and experiment, which stays at a factor of $\approx 3$.

Furthermore, the strength of the short-range electron-electron interaction depends on the QD's ground state orbital wave function in real space, and the corresponding prefactor can be approximated as \cite{Knothe2020JunQuartetStates, Moller2021Dec, Knothe2022AprBatmanTheo}
\begin{equation}
\mathcal{J} \approx \int d\mathbf{r} \; |\Psi(\boldsymbol{r} )|^2 |\Psi(\boldsymbol{r} )|^2.
\end{equation}
Assuming a coupling constant $|g_\perp| = 0.16\, \mathrm{eVnm}^2$ \footnote{$g_\perp$ is defined as in Refs.~\cite{Knothe2020JunQuartetStates, Moller2021Dec, Knothe2022AprBatmanTheo}}, which is in the same order of magnitude as in \cite{Moller2021Dec}, we achieve a good agreement with the experimental data.

\subsection{Valley g-factor depends on QD occupation}
\label{app:g_v_N}
For determining the valley g-factor as a function of N, we first extract $\Delta \mu_N \, (B_\perp)$ from the outline of the conducting region of the finite bias data presented in Fig.~\ref{f3}, as highlighted by the colored lines. The result (for $B_\perp > 0)$ can be seen in Fig.~\ref{f4} (b). In principle, we could also utilize $\Delta \mu_N (B_\perp)$ from the data presented in Fig.~\ref{f2}(a,b), but due to a better signal-to-noise ratio, we take the data from Fig.~\ref{f3}for $N \leq 8$ and only use the data from Fig.~\ref{f2}(b) for $ 8 < N \leq 12$.
Next, we perform linear fits on $\Delta \mu_N (B_\perp)$ for extracting the g-factor.
For the first and forth transition we fit in the range of  $ |B_\perp| \lessapprox  0.85~T$, depending on where the crossing with the transition from the next orbitals happens. For the second and third transition we fit in the range of $ B_\mathrm{TS} \lessapprox |B_\perp| \lessapprox  0.85~T$.  
Utilizing equations (1-5), we can extract the g-factors from the obtained slopes according to
\begin{align}
    g^{(1)}_v &= 2 \frac{\alpha_1}{\mu_B} - g_s \;,\\
    g^{(2)}_v &= 2 \frac{\alpha_2}{\mu_B} + \frac{1}{2}  (g_s + g^{(1)}_v) \; , \\
    g^{(3)}_v &= 2 \frac{\alpha_3}{\mu_B} - g_s - 2 g^{(2)}_v , \\
    \tilde{g}^{(3)}_v &= 2 \frac{\alpha_4}{\mu_B} + g_s \;,
\end{align}
with the fitted slope of each charge transition $\alpha_N$ (see grey dashed lines in Fig.~\ref{f2}(b)). As already apparent from equation (1-5), the valley g-factor $g^{(4)}_v$ does not contribute to the slopes as the shell is completely full. Therefore, the forth charge transition also yields the g-factor for $N = 3$, which we label $\tilde{g}^{(3)}_v$. The difference to $g^{(3)}_v$ is due to the increased FG voltage at the forth charge transition.

With this, we take a closer look at the modulation of the valley g-factor within one shell, as visible in Fig.~\ref{f2}(c), which we attribute to mainly two effects. First, the g-factor decreases as soon as more than one electron occupies the same shell. In a simple picture, we assume this is due to the electrons repelling each other, widening the wavefunction of the QD compared to the wavefunction of only one electron in a shell. In agreement with the trend observed in Fig.~\ref{f5}, a larger wavefunction in real space leads to smaller g-factors. On the other hand, for each additional electron within a shell, the FG voltage needs to be increased by $\Delta V_\mathrm{FG} \approx 0.1\,$V, which slightly changes the confinement potential and therefore the g-factor. The magnitude of the latter effect can be estimated from the third and fourth data point in each shell in Fig.~\ref{f2}(c), as they both correspond to the valley g-factor of the shell having three electrons, but are evaluated at different FG voltages ($g^{(3)}_v$ and $\tilde{g}^{(3)}_v$). One finds an increase of the g-factor by $\Delta g_v / \Delta V_{FG} \approx 1 / (100 \, \mathrm{mV})$ within a single shell due to the change in FG voltage.\\

\subsection{Multi-particle excited state transitions}
\label{app:ES_transitions}
The finite bias spectroscopy data presented in Fig.~\ref{f3} shows features in the differential transconductance, $dI/dV_\mathrm{FG}$, which shift in an out-of-plane magnetic field. They can be explained in terms of transitions between $N$ and $N+1$ particle states, which enter or leave the bias window depending on the magnetic-field dependent energy difference between the involved states~\cite{Moller2021Dec, Knothe2022AprBatmanTheo}. In the following, we will discuss which transitions give rise to the labeled features and explain their magnetic field dependency. For understanding the slopes of each transition, we only focus on the valley Zeeman effect, as it is much larger than the spin Zeeman effect, which is too small to be resolved in the data presented in Fig.~\ref{f3}. The slope is thus only given by the difference of the effective valley g-factors of the two states involved in one transition. For example, feature \textbf{a} is due to a transition from an excited single particle state to the two particle GS, $\ket{K^+ \uparrow \downarrow} \rightarrow \ket{T^s_{0, \pm} \, S^v}$. The two-particle state is a valley singlet, thus having an effective valley g-factor of zero, while the effective valley g-factor of the single particle state $\ket{K^+ \uparrow \downarrow}$ is given by $+g^{(1)}_v$, increasing the energy of the state with $B_\perp$. Thus, the chemical potential necessary to allow this transition reduces, leading to a negative slope of \textbf{a}. The slopes of the following transitions can be understood similarly, comparing the effective valley g-factors of the initial and final states. For better readability, we will omit the spin quantum number from now on.  As already discussed in the caption of Fig.~\ref{f3}, \textbf{b} is a transition from the single-particle GS to the valley singlet two particle state $\ket{K^-} \rightarrow \ket{S^v}$, which has a positive slope due to the single-particle state losing energy with $B_\perp$. Transition \textbf{c} involves the states $ \ket{T^v_-} \rightarrow \ket{K^-\, K^+ \, K^-}_a$, which becomes the ground state to ground state (GS to GS) transition for $B > B_\mathrm{TS}$. For lower magnetic fields, it requires less chemical potential than the GS to GS transition. Thus, it leads to a decrease of the conductance when it leaves the bias window, which shows as negativ transconductance at \textbf{c}. Transition \textbf{d} involves the states $\ket{S^v} \rightarrow \ket{K^-\, K^+ \, K^+}_a$, with the latter being an excited state of the (asymmetric) three particle states. Transition \textbf{e} involves the states  $\ket{K^-\, K^-\, (K^-)_2}_a \rightarrow \ket{K^-\, K^-\, (K^-)_2  \, K^+}_a$, where the subscript 2 refers to the fact that the electron is in the next higher orbital. It thus involves an exited state of both the three and four particle states. Transition \textbf{f} involves  $\ket{K^-\, K^+\, K^- \, (K^-)_2}_a \rightarrow \ket{\mathrm{full~shell}\, \,(K^-)_2}_a$, i.e. a transition from an excited four particle state, where one of the four electrons already occupies the next higher orbital, to the five particle GS. Transition \textbf{g} involves $\ket{K^- \, K^- \, K^+, \, (K^-)_2 \, (K^-)_2} \rightarrow \ket{\mathrm{full} \, \mathrm{shell} \, (K^-)_2 \, (K^-)_2}$. All primed and unprimed transitions, i.e. \textbf{a} $\leftrightarrow$ \textbf{a'}, \textbf{b} $\leftrightarrow$ \textbf{b'}, etc., can be explained in the same way, as they have the same amount of electrons in the highest unpaired shell.

\subsection{Data availability}
The data and evaluation scripts supporting the findings of this work are available in a Zenodo repository under XXX.

\bibliography{literature}
\end{document}